\documentclass[twocolumn,showpacs,preprintnumbers,amsmath,amssymb,superscriptaddress]{revtex4}
\usepackage{amssymb}
\usepackage{graphicx,color}% Include figure files
\usepackage{bm}% bold math

\begin{document}

\title{Nuclear modification of high-$p_{T}$ hadron spectra in $p+A$ collisions at LHC}

\author{Rong Xu}
\affiliation{Key Laboratory of Quark and Lepton Physics (MOE) and Institute of Particle Physics, Central China Normal University, Wuhan 430079, China}

\author{Wei-Tian Deng}
\affiliation{Frankfurt Institute for Advanced Studies,
Ruth-Moufang-Strasse 1, D-60438 Frankfurt am Main, Germany}

\author{Xin-Nian Wang}
\affiliation{Institute of Particle Physics, Central China  Normal University, Wuhan 430079, China}
\affiliation{Nuclear Science Division, MS 70R0319,
Lawrence Berkeley National Laboratory, Berkeley, California 94720}

\begin{abstract}
Multiple parton scatterings in high-energy $p+A$ collisions involve multi-parton correlation inside the projectile 
and color coherence of multiple jets which will lead to nuclear modification of final hadron spectra relative 
to that in $p+p$ collisions.  Such modifications of final hadron spectra in $p+A$ collisions are
studied within the HIJING 2.1 model. Besides parton shadowing and transverse momentum broadening, which
influence final hadron spectra at intermediate $p_{T}$,  modification of parton flavor composition due to
multiple scattering and hadronization of many-parton jets system are shown to lead to suppression of  final 
hadrons at large $p_{T}$ in $p+A$ collisions at  the Large Hadron Collider. 

\end{abstract}


\maketitle

In search for the formation of quark-gluon plasma (QGP) in high-energy heavy-ion collisions, 
jet quenching or suppression of high-$p_{T}$ hadrons \cite{Wang:1991xy} can be used for tomographic study of medium 
properties \cite{Gyulassy:2001nm,Wang:2002ri}. However, initial multiple parton scatterings in 
cold nuclei before the formation of  QGP can also lead to nuclear modification of final high-$p_{T}$ 
hadron spectra that needs to be understood for a quantitative study of QGP properties through jet quenching.

Parton shadowing or nuclear modification of parton distributions inside large nuclei is one of the most
studied cold nuclear effects. The QCD evolution, however, dramatically reduces the effect on
final high-$p_{T}$ hadron spectra in $p+A$ collisions  \cite{Eskola:1998df,Vitev:2002pf,QuirogaArias:2010sg,Barnafoldi:2011px}. 
Modification of parton distributions in the projectile nucleon through parton energy loss \cite{Neufeld:2010dz,Xing:2011fb} is 
estimated to have a small effect on hadron spectra in the central rapidity region of high-energy $p+A$ collisions. 
These cold nuclear effects have also been studied within the
Color Glass Condensate (CGC) model \cite{McLerran:1993ni,Kharzeev:2003wz,JalilianMarian:2011dt}. 
However, multiple parton correlations inside the projectile nucleon should also be considered.
 Energy-momentum and valence quark number conservation alone could modify the momentum and
flavor dependence of final-state parton and hadron spectra at large $p_{T}$. Hadronization of multiple-jet systems
from multiple scattering in $p+A$ collisions could also affect final hadron spectra at intermediate and large $p_{T}$.  

In this paper, we study the nuclear modification of parton and hadron spectra
due to parton shadowing, initial and final-state transverse momentum broadening, 
final-state parton flavor composition and hadronization of multiple jets within the
HIJING  \cite{Wang:1991hta} Monte Carlo model. It is
based on a two-component model for hadron 
production in high-energy nucleon and nuclear collisions. The soft and hard components are 
 separated by a cut-off $p_0$ in the transverse momentum exchange. Hard parton
 scatterings with  $p_T>p_0$ are assumed to be described by the perturbative QCD (pQCD) and 
soft interactions are approximated by string excitations with an effective cross section $\sigma_{\rm soft}$.
In HIJING 2.0 \cite{Deng:2010mv}, parton distribution functions (PDF's) are given by the 
Gluck-Reya-Vogt(GRV) parameterization \cite{Gluck:1994uf}. An energy-dependent  cut-off $p_0(\sqrt{s})$ 
and soft cross  section $\sigma_{\rm soft}(\sqrt{s})$ are assumed in order to fit
experimental values of the total and inelastic cross sections of $p+p/(\bar p)$ collisions and the energy-dependence 
of the central rapidity density of charged hadron multiplicities. HIJING 2.0 can describe most features of hadron 
production in $p+p$ collisions at colliding energies up to 7 TeV at the Large Hadron Collider (LHC) \cite{Deng:2010mv}.

In $p+A$ and $A+A$ collisions, one has to consider nuclear modification of initial parton distributions
due to multiple scatterings. HIJING 2.0 employes a factorized form of parton distributions in nuclei,
\begin{equation}
f_{a/A}(x,Q^2,b)=AR_a^A(x,Q^2,b)f_{a/N}(x,Q^2),
\label{eq:shdw}
\end{equation}
where $Q$ is the momentum scale, $b$ the impact-parameter, $x$ the momentum fraction
carried by the parton $a$ and $R_a^A(x,Q^2,b)$ is the impact-parameter-dependent nuclear 
modification factor as given by the new HIJING parameterization \cite{Li:2001xa}. 
It was fit to the centrality dependence of the pseudo-rapidity multiplicity density per 
participant  pair $2 dN_{ch}/d\eta/N_{\rm part}$ at the Relativistic Heavy-ion Collider (RHIC), and also 
describes the data at the LHC energies as well \cite{Deng:2010mv}.

In addition to modification of initial parton distributions, multiple scatterings inside a nucleus should 
also lead to transverse momentum ($k_{T}$) broadening of both initial and final-state partons
The $k_{T}$-broadening is responsible for the Cronin effect or enhancement of intermediate-$p_{T}$ 
hadron spectra in $p+A$ collisions \cite{Cronin:1974zm}.  It is implemented in HIJING 2.1
as a $k_{T}$-kick by each binary nucleon-nucleon scattering to the initial partons of a hard
scattering as well as final-state partons produced prior to the current scattering. The $k_{T}$-kick
for each scattering follows a Gaussian distribution and an energy dependence of the width,
\begin{equation}
\langle k_{T}^{2}\rangle = 0.14\log(\sqrt{s}/{\rm GeV})-0.43 \;\;\; {\rm GeV}^{2},
\end{equation}
is assumed to describe the measured Cronin effect in $p+A$ collisions. After the $k_{T}$-kick,
longitudinal momenta of partons are reshuffled pair-wise between projectile and
target partons in order to ensure four-momentum conservation. Consequently, introduction
of the $k_{T}$-kick will also influence hadron rapidity distributions. We have to fine-tune
the gluon shadowing parameter $s_{g}$ \cite{Deng:2010mv,Li:2001xa} again to describe experimental data on rapidity 
distributions.

\begin{figure}
  \centering
\includegraphics[width=0.40\textwidth]{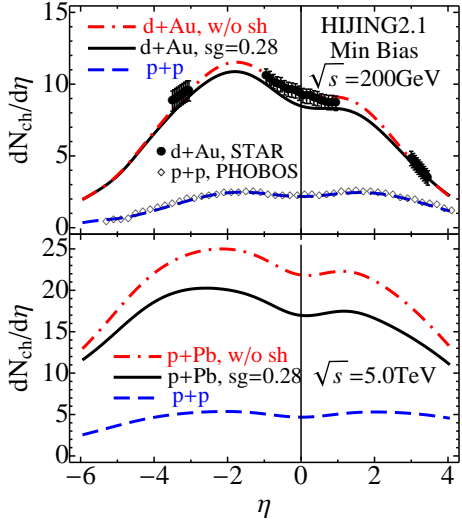}

  \caption{(color online) Charged hadron rapidity distribution from HIJING 2.1 with (solid)
and without (dot-dashed) parton shadowing in $p(d)+A$ and $p+p$ (dashed) 
collisions at $\sqrt{s}_{NN}=200$ GeV (upper) and 5.0 TeV (lower) as compared
to data from the STAR \cite{Abelev:2007nt} and PHOBOS \cite{Alver:2010ck} experiment.}
  \label{fig:dndy}
\end{figure}

Shown in Fig.~\ref{fig:dndy} are HIJING 2.1 results on rapidity distributions of charged hadrons in 
minimum-biased $d+Au$ (upper panel) and $p+Pb$ (lower panel) collisions (solid)
at $\sqrt{s}_{NN}=200$ GeV and 5.0 TeV, respectively,  as compared to $d+Au$ data 
from the STAR experiment \cite{Abelev:2007nt}. Here the gluon shadowing parameter is set to $s_{g}=0.28$.
Also shown are HIJING2.1 results for $p+p$ collisions (dashed) as compared to the PHOBOS data \cite{Alver:2010ck} 
and $p(d)+A$ collisions without parton shadowing (dot-dashed).  Gluon shadowing in HIJING 2.1
is seen to suppress the hadron yield in mid and forward rapidity region, especially at the LHC energy.
Measurement of the hadron rapidity distribution at LHC therefore would provide further constraints on 
the gluon shadowing.

The hard processes in HIJING are based on lowest order pQCD in which the single jet inclusive cross section,
\begin{eqnarray}
\frac{d\sigma^{jet}_{pA}}{dy_{1}d^{2}p_{T}}=K \int dy_{2}d^{2}b t_{A}(b)\sum_{a,b,c} x_{1}f_{a/p}(x_{1},p_{T}^{2}) \nonumber \\ 
\times x_{2}f_{a/A}(x_{2},p_{T}^{2},b)\frac{d\sigma_{ab\rightarrow cb }}{dt},
\end{eqnarray}
is directly proportional to nuclear parton distributions $f_{a/A}(x_{2},Q^{2},b)$. 
Here $x_{1,2}=p_{T}(e^{\pm y_{1}}+e^{\pm y_{2}})/\sqrt{s}$ are
fractional momenta of the initial partons, $y_{1,2}$ are rapidities of the final-state parton jets 
and $K\approx 2$ accounts for higher-order corrections. The
nuclear thickness function is normalized to $\int d^{2}b t_{A}(b)=1$. In the central rapidity region ($y_{1,2}=0$), 
transverse momenta of final-state jets determine the momentum fractions $x_{1,2}=2p_{T}/\sqrt{s}$ 
of the initial colliding partons.  Therefore, $p_{T}$ spectra of produced partons via hard scatterings 
in $p+A$ collisions should depend on the nuclear modification of parton distributions in nuclei. 
To study parton shadowing and other cold nuclear effects, we calculate the nuclear modification factor 
for final-state parton and hadron $p_{T}$ spectra,
\begin{equation}
R_{pA}(p_{T})=\frac{dN_{pA}/dyd^{2}p_{T}}{\langle N_{\rm bin}\rangle dN_{pp}/dyd^{2}p_{T}},
\end{equation}
where $\langle N_{\rm bin}\rangle$ is the average number of binary nucleon-nucleon interactions in $p+A$ 
collisions.

\begin{figure}
  \centering
\includegraphics[width=0.4\textwidth]{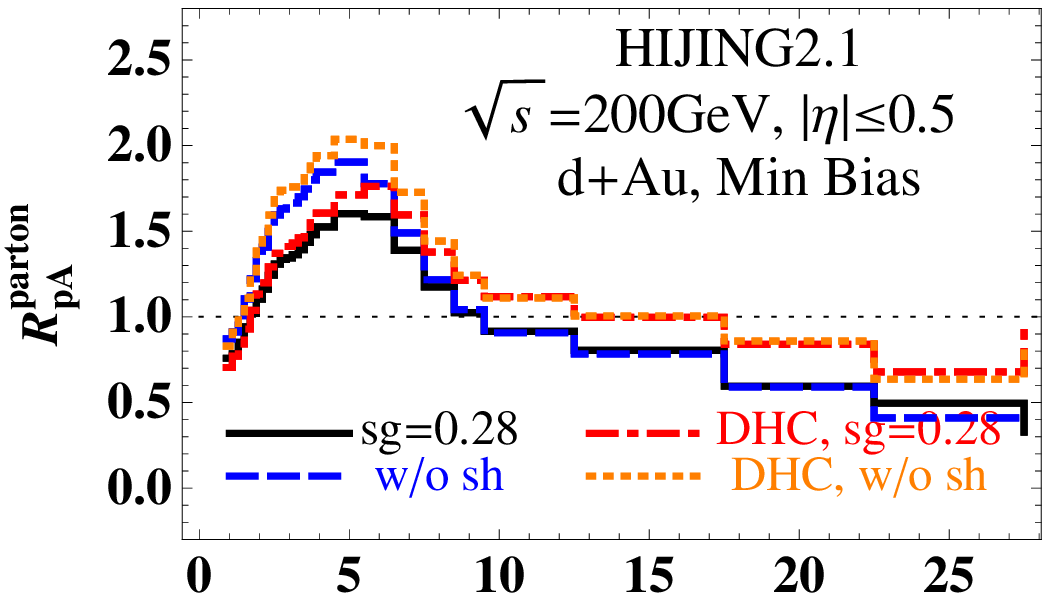}
\vspace{-0.2in}

\hspace{-2pt}\includegraphics[width=0.4\textwidth]{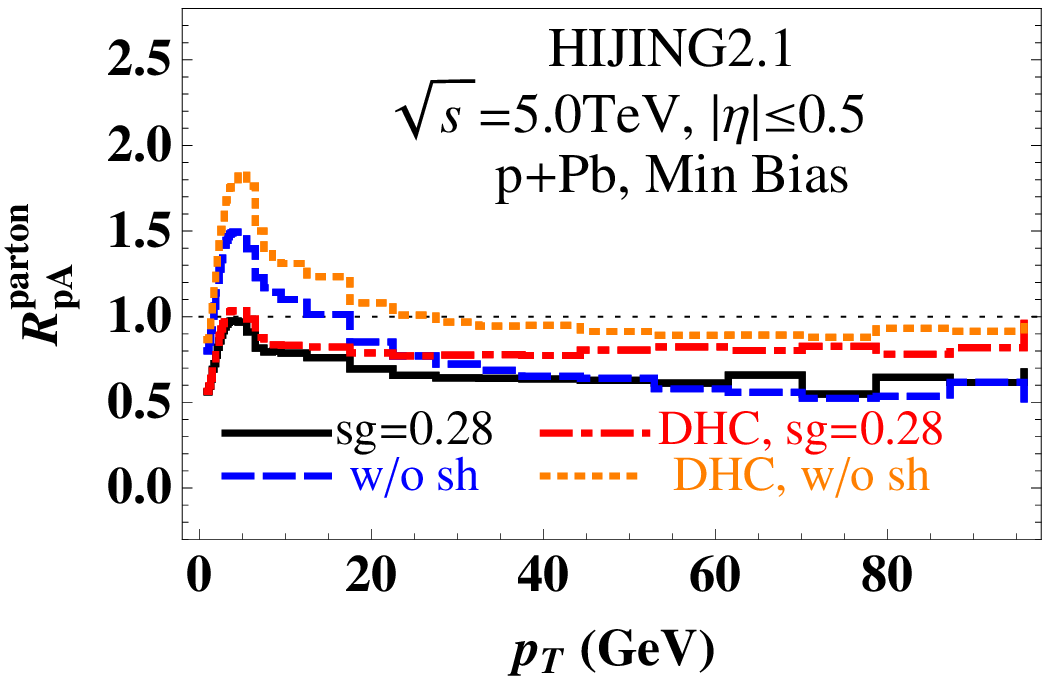}

  \caption{(color online) Nuclear modification factor for parton $p_{T}$ spectra in $p(d)+A$
  collisions at $\sqrt{s}_{NN}=200$ GeV (upper) and 5.0 TeV (lower) from HIJING 2.1 with (solid), 
  without (dashed) parton shadowing and with de-coherent hard scatterings (DHC) (dot-dashed and doted).}
  \label{rparton}
\end{figure}

Shown in Fig.~\ref{rparton} are parton modification factors from the HIJING 2.1 for minimum-biased $d+Au$ and $p+Pb$
collisions at $\sqrt{s}_{NN}=200$ GeV and 5.0 TeV, respectively. The enhancement of parton spectra at 
intermediate $p_{T}$ is the signature effect of $k_{T}$-broadening, which should disappear at
large $p_{T}$. In the HIJING model, hard scatterings are simulated each time the projectile parton interact with 
a target parton alongside soft interactions with the remnants of colliding nucleons. HIJING ensures 
energy-momentum conservation by subtracting energy-momentum transfers in the previous hard and soft
interactions from the projectile and thus restricting energy available for subsequent hard collisions. 
Since the time scale of hard scatterings is much shorter than soft interactions, such a coupling between
hard and soft interactions might not be physical. As an option in our study, we turn off such a coupling
and do not restrict energy available for hard scatterings. We denote the results as DHC (de-coherent hard scattering). 
One can see from Fig.~\ref{rparton}, both parton shadowing and soft-hard coupling 
suppresses high-$p_{T}$ parton production. These features of parton spectra will be reflected in the 
final hadron spectra after hadronization.

%Without shadowing and soft-hard coupling, 
%the parton spectra in $p+A$  should be exactly proportional to the number of binary collisions
%and the nuclear modification factor approaches to 1 at large $p_{T}$. 

\begin{figure}
  \centering
\includegraphics[width=0.4\textwidth]{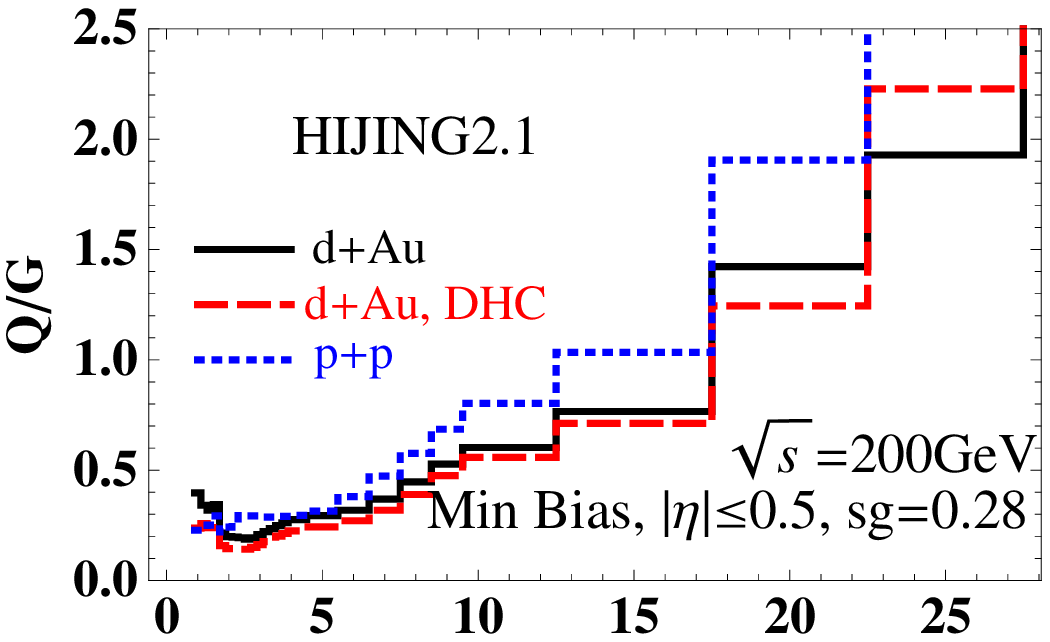}
\vspace{-0.2in}

\hspace{-2pt}\includegraphics[width=0.4\textwidth]{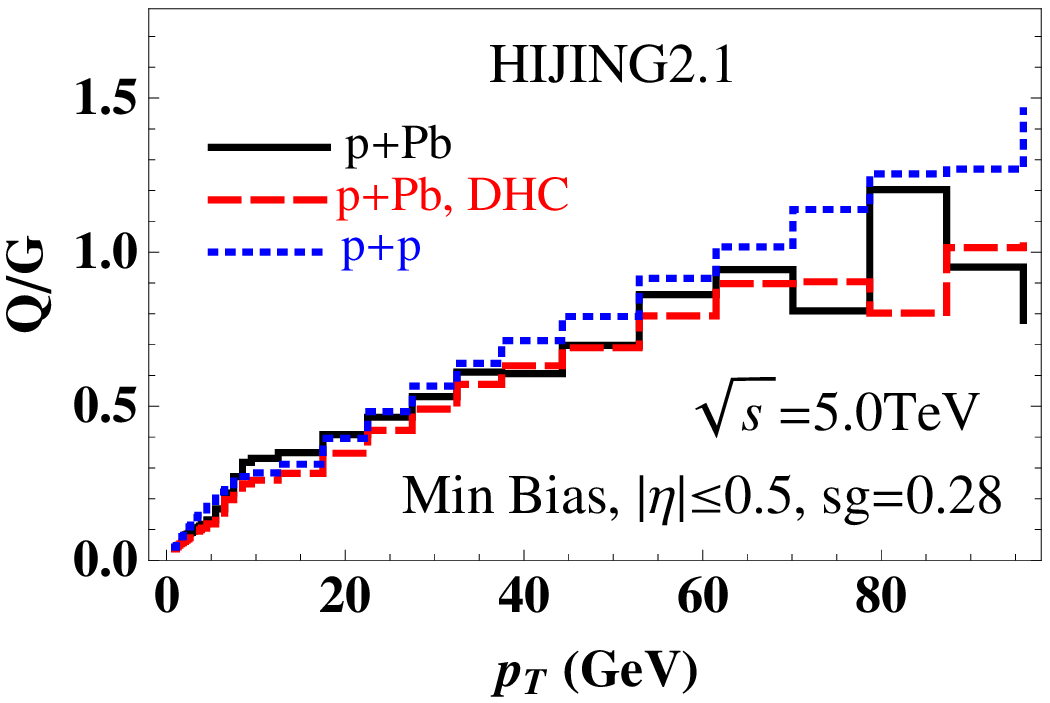}

  \caption{(color online) Quark to gluon ratio as a function of $p_{T}$ in $p+p$ (dotted) $p(d)+A$  collisions 
  at the RHIC (upper) and LHC (lower) energies from HIJING 2.1 without (solid) and with (dashed)
  de-coherent hard scatterings (DHC).}
  \label{qgratio}
\end{figure}

\begin{figure}
  \centering
  \includegraphics[width=0.4\textwidth]{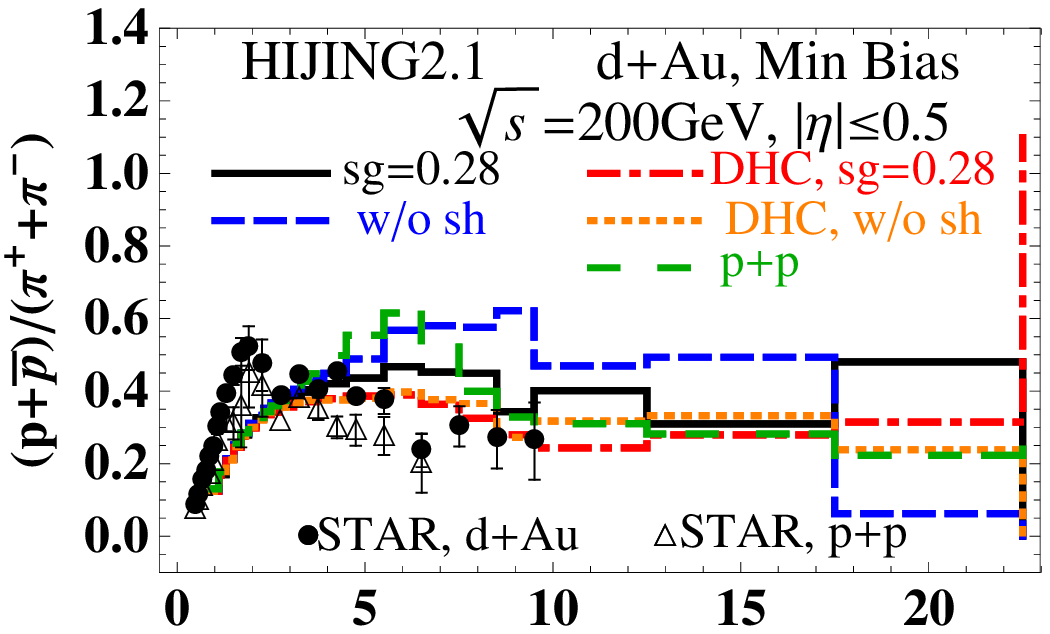}
  
    \vspace{-0.2in}
\includegraphics[width=0.4\textwidth]{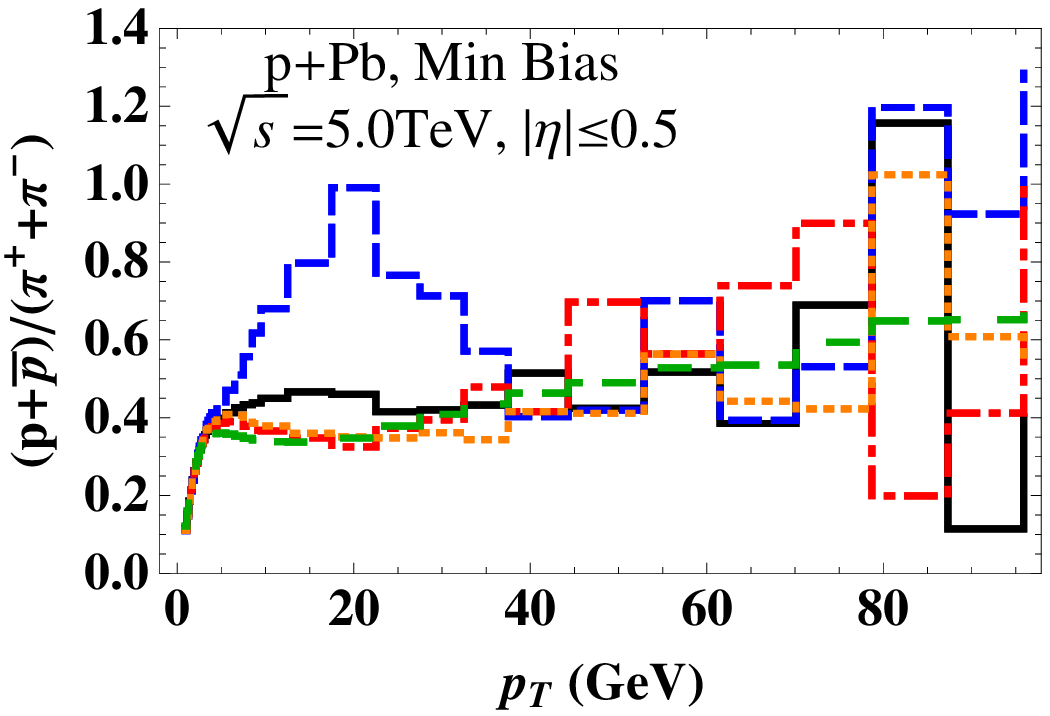}
 \caption{(color online) Proton to pion ratio at RHIC and LHC from HIJING 2.1
 with different options: default (solid), without shadowing (ashed), DHC (dot-dashed), DHC without parton 
  shadowing (dotted) in $d+Au$ collisions at RHIC and $p+Pb$ at LHC. The ratio in $p+p$ collisions is
  also shown (long dashed). The RHIC experimental data are from STAR \cite{Adams:2006nd}. }
  \label{p-pi-ratio}
\end{figure}

In $p+A$ collisions, the projectile proton scatters with multiple nucleons inside the target, each
 with finite probability of hard parton scattering. The leading contribution comes from independent hard 
 scatterings, each involving independent initial partons from projectile and target nucleons.   
 In principle, a fast projectile parton can also suffer multiple hard scatterings which will affect the final-state
 parton spectra but do not increase the projectile's contribution to the final-state parton multiplicity.  
 Momentum correlation of  multiple partons inside the projectile is neglected beyond the conservation 
 of the total momentum.  Flavor conservation, however, will limit the availability 
  of valence quarks from the projectile for each of these multiple independent hard scatterings 
  as the projectile plunges through the target nucleus. Such limitation on the number of valence
  quarks available for each independent hard scattering will lead to a change in the flavor composition of
  produced partons per average binary nucleon-nucleon collision. Shown in Fig.~\ref{qgratio} are the ratios
  of produced quark and gluon $p_{T}$ spectra  in the central rapidity region of $d+Au$ and $p+Pb$
  collisions at $\sqrt{s}_{NN}=200$ GeV and 5.0 TeV, respectively, as compared to that in $p+p$ collisions.
  The quark number conservation in multiple scatterings is seen to lead to smaller quark/gluon ratio in $p+A$
  relative to $p+p$ collisions. The suppressed quark yield and high quark/gluon ratio, especially at RHIC,  
  will lead to the modification of final-state parton spectra at large $p_{T}$ as shown in Fig.~\ref{rparton}. 
  Because gluon fragmentation functions are softer than quarks, the increased fraction of gluon jets 
  in $p+A$ collisions will lead to apparent suppression of high $p_{T}$ hadron spectra relative to $p+p$ collisions.
  
The modified flavor composition of the final parton due to multiple scattering in pA collisions could also change the
flavor composition of final hadrons due to flavor-dependence of the fragmentation functions. Gluon jets in general
have higher proton-to-pion ratio than in quark jets. HIJING has a strong gluon shadowing which suppresses larger $p_{T}$ gluon jets. 
This should reduce the proton-to-pion ratio in HIJING 2.1 results as compared to the results without gluon shadowing 
as shown in Fig.~\ref{p-pi-ratio}.  This reduction of proton-to-pion ratio by gluon shadowing is much stronger at the LHC energy. 
Nevertheless, the proton-to-pion ratio in $d+Au$ collisions at RHIC and $p+Pb$ collisions at LHC will still be larger than 
that in $p+p$ collisions because of the enhanced gluon-to-quark ratio in $p+A$ collisions relative to that in $p+p$ as shown
in Fig.~\ref{qgratio}. Such a trend is already seen in the STAR data \cite{Adams:2006nd} at RHIC. 
The flavor composition of final hadrons in $p+A$ collisions could also be influenced by other nonperturbative 
mechanisms such as parton recombination during hadronization \cite{hwa} and hadronic final-state rescattering \cite{Kopeliovich}.

Another process that can also modify final hadron spectra in $p+A$ collisions is jet 
hadronization or fragmentation. HIJING follows the same routine as in PYTHIA that groups jet partons together
with project and target remnants to form color-singlet string systems. In HIJING, jet parton showers 
from all hard scatterings, including initial and final-state radiations, are ordered in rapidity and  gluons are then 
connected to the valence quark and diquark of the associated projectile or target nucleons as kinks to form 
string systems. These string systems are hadronized according to the Lund string model \cite{Andersson:1983ia}.
The projectile can undergo multiple scatterings in $p+A$ collisions. Its string system therefore has more gluons 
attached as compared to that in $p+p$ collisions. Hadrons
from such a string system are expected to be softer than independent fragmentation of individual gluons.
To test the sensitivity of nuclear modification of final hadron spectra to jet hadronization, we also
consider the independent fragmentation mechanism in HIJING for both $p+p$ and $p+A$ collisions.

 \begin{figure}
  \centering
\includegraphics[width=0.4\textwidth]{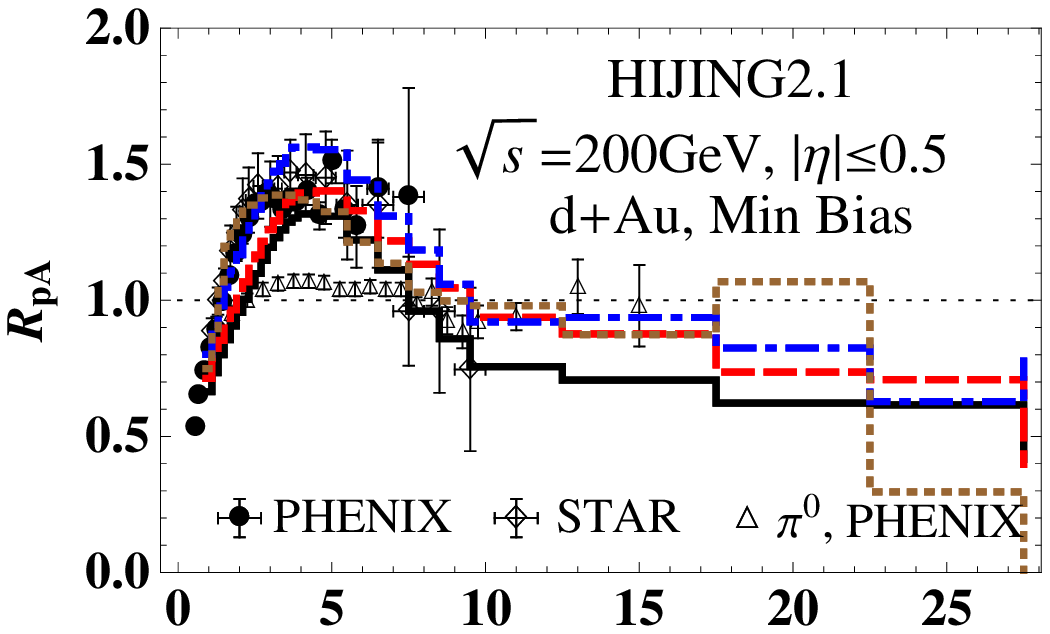}
\vspace{-0.2in}
\hspace{-3pt}\includegraphics[width=0.4\textwidth]{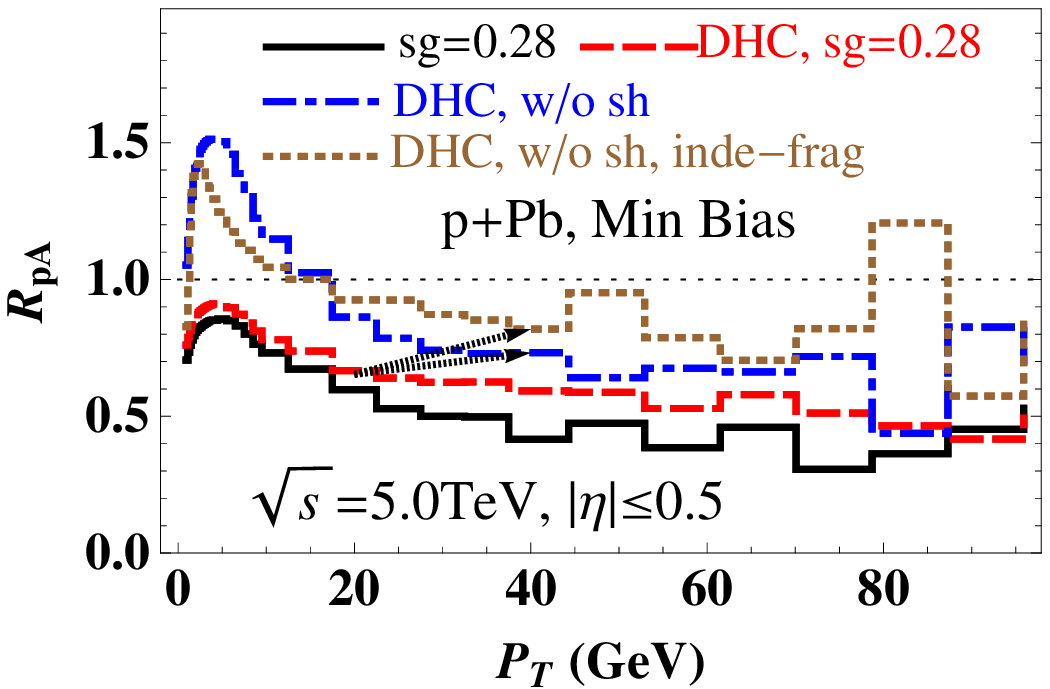}

  \caption{(color online) Nuclear modification factor for charged hadrons in 
 $p+A$ collisions from HIJING 2.1 with different options: default (solid), DHC (dashed), DHC without parton 
  shadowing (dot-dashed) and DHC without parton shadowing and independent fragmentation (dotted),
as compared to data from the PHENIX and STAR experiment \cite{Adler:2003ii,Adams:2003im}. 
The arrowed lines indicate the most possible behavior of the nuclear modification factors at LHC between
low and high $p_{T}$ regions.}
  \label{rhadron}
\end{figure}

Shown in Fig.~\ref{rhadron} are nuclear modification factors for charged hadrons in $d+Au$ and
$p+Pb$ collisions at $\sqrt{s}_{NN}=200$ GeV and 5.0 TeV, respectively, as compared to existing data
from both the PHENIX and STAR expok, eriment at RHIC \cite{Adler:2003ii,Adams:2003im}. At low $p_{T}$, 
hadron production is dominated by hadronization of soft strings which is proportional
to the number of participant nucleons $N_{\rm part}=1+N_{\rm bin}$. The nuclear modification factor
should approach $R_{pA}\approx (1+1/N_{\rm bin})/2 \approx 1/2$ in the limit $N_{\rm bin}\gg1$ .
At large $p_{T}$, incoherent hard parton scatterings become dominate and $R_{pA}\approx 1$ 
without any nuclear modification. The peak in the nuclear modification factor is generally known as the 
Cronin enhancement of hadron spectra at intermediate $p_{T}$ due to $k_{T}$-broadening through multiple scatterings.
Such a peak has been observed in pA collisions at the Fermilab energies ($E_{\rm lab}=200-800$ GeV) \cite{Wang:1998ww}
 and the position of the peak at around 4 GeV/$c$ has no apparent energy dependence and remains the same
 in HIJING results at both RHIC and LHC energies.  Nuclear shadowing of parton distributions at LHC will only 
 affect the overall value of $R_{pA}$ at intermediate $p_{T}$, but the peak-feature and its position remain the same.
 This is a unique prediction of the $k_{T}$-broadening due to multiple scattering as implemented in HIJING2.1. It is
 in sharp contrast to the prediction of disappearance of the peak in gluon saturation model or relocation to much higher
 $p_{T}$ due to strong anti-shadowing in the collinear pQCD parton model.  Experimental test of these features will 
 shed light on the parton dynamics in high-energy nuclei.

At very large $p_{T}$, parton shadowing in nuclei, soft-hard coupling and enhanced gluonic jets due to 
valence quark conservation all lead to suppression of charged hadron spectra.
Fragmentation of the projectile string system with multiple gluon jets in $p+A$
collisions seems to further suppress high-$p_{T}$ hadron spectra as compared to the independent
fragmentation (dotted).  These nuclear effects are consistent with existing $d+Au$ data at RHIC. 
However, $p+Pb$ collisions at LHC will be able to test these scenarios within the available $p_{T}$ range. 
Since parton shadowing will disappear at large $p_{T}$ \cite{Eskola:1998df} due to
QCD evolution and hard scatterings are de-coherent from soft interactions, the 
nuclear modification factor at LHC will likely follow the default and DHC results at low $p_{T}$ and approach DHC 
without shadowing at large $p_{T}$  with possible further modifications due to hadronization of multiple jets.
In Fig.~\ref{rhadron}, we indicate this trend by joining these two set of results with arrowed lines at intermediate $p_{T}$

\begin{figure}
  \centering
\includegraphics[width=0.4\textwidth]{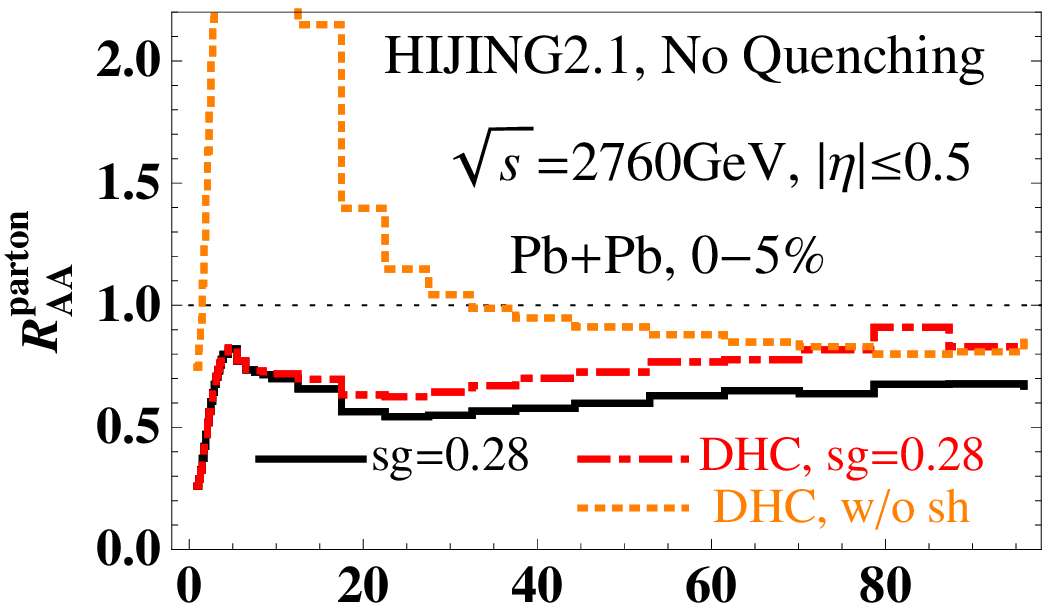}
\vspace{-0.2in}

\hspace{-3pt}\includegraphics[width=0.4\textwidth]{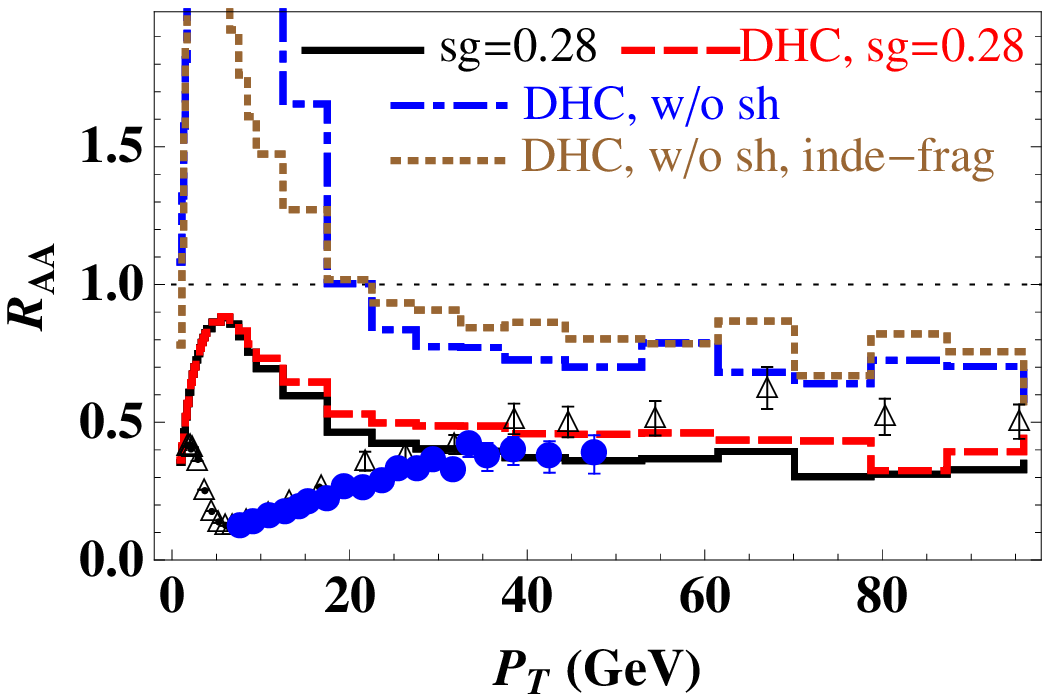}

  \caption{(color online) Nuclear modification factors for final parton (upper panel) and charged hadrons (lower panel) in 
 central Pb+Pb collisions at $\sqrt{s}=2.76$ TeV from HIJING 2.1 with different options: default (solid), DHC (dashed), 
 DHC without parton shadowing (dot-dashed) and DHC without parton shadowing and independent fragmentation (dotted),
as compared to data from the ALICE and CMS experiment \cite{Aamodt:2010jd,CMS:2012aa}. Jet quenching due to final-state interaction
is not considered in the HIJING calculations.}
  \label{raa}
\end{figure}

The initial-state interaction and fragmentation of multiple jet systems should also affect the final hadron spectra
in high-energy heavy-ion collisions in addition to jet quenching due to final-state interaction. The existing
experimental data on hadron spectra modification should provide some constraints on the initial-state parton interactions
and the final hadronization mechanism. Shown in Fig.~\ref{raa} are nuclear modification factors for final
parton (upper panel) and hadron spectra in central $Pb+Pb$ collisions  at $\sqrt{s}=2.76$ TeV. Note that there is
no jet quenching effect in the calculation shown. On the parton level, the nuclear modification at large $p_{T}$ mainly 
comes from parton shadowing which should be small if the scale dependence of shadowing is taken into account.
The final hadron spectra at large $p_{T}$ are however significantly suppressed due to the Lund string fragmentation of
multiple parton systems (solid and dashed curves in the lower panel) leaving no room for additional suppression by
jet quenching as compared to experimental data from ALICE \cite{Aamodt:2010jd} and CMS \cite{CMS:2012aa}.
This is a strong indication that the Lund string fragmentation for hadronization is no longer a valid mode for 
multiple gluonic jet systems in heavy-ion collisions where a deconfined quark-gluon plasma is formed. 
Recent studies show that multiple parton scatterings inside a quark-gluon plasma can in fact change the
the color flow of the jet parton shower from that in 
vacuum \cite{Leonidov:2010he,CasalderreySolana:2011rz,MehtarTani:2010ma,Beraudo:2012bq}. One therefore
should expect the fragmentation processes in heavy-ion collisions to be significantly modified from that in vacuum.
An independent jet fragmentation mechanism such
as parton recombination model might be a better alternative where local parton phase distributions determine the final hadron
spectra and global color flow of the multiple jet system is no longer important. Since multiple gluonic jet systems in pA collisions
can be considered as precursors of the quark-gluon plasma, investigation of the final hadron spectra at large $p_{T}$ in
$p+A$ collisions can shed light on the hadronization mechanism of energetic jets after they escape jet quenching  and emerge from the quark-gluon plasma in high-energy heavy-ion collisions.

In summary, we have studied the nuclear modification of hadron spectra in $p+A$ collisions
at both the RHIC and LHC energies within HIJING2.1 Monte Carlo model and their sensitivities
to cold nuclear effects. Multiple parton correlations in the projectile due to valence quark number conservation, 
string fragmentation of multiple gluon jets as well
as parton shadowing can all lead to suppression of high-$p_{T}$ hadron spectra in $p+A$
relative to $p+p$ collisions. Comparisons between HIJING2.1 results without jet quenching and existing
data on large $p_{T}$ hadron suppression at LHC indicate an independence fragmentation mechanism for
multiple jet systems such as in pA and AA collisions.  Experimental test of these effects in $p+Pb$
collisions at LHC and its implications for hard scatterings in $A+A$ collisions is crucial 
for the study of jet quenching and extraction of jet transport properties in the hot quark-gluon plasma.

\section*{Acknowledgement}

We thank M. Gyulassy for helpful discussions. This work was supported in part by the NSFC under the 
project  No. 10825523 and 11221504,  by self-determined research funds of CCNU from the 
colleges' basic research and operation of MOE,
Helmholtz International Center for FAIR within the framework of the LOEWE program 
launched by the State of Hesse, and US Depart of Energy under Contract No. DE-AC02-05CH11231 
and within the framework of the JET Collaboration.

\end{document}